\documentclass[eqsecnum,superscriptaddress,nofootinbib,11pt,showkeys]{revtex4}
\usepackage{amsmath}
   \usepackage{bm}
\usepackage{amsthm,amscd}
\usepackage{mathrsfs}
\usepackage{verbatim}
\usepackage{amsfonts,amsmath,latexsym,amssymb,txfonts,bm}
\usepackage[american]{babel}
\usepackage{dsfont}
\usepackage[dvips]{graphicx}
\usepackage{latexsym}
\usepackage{epsfig}
\newcommand{\diff}{\mathrm{d}}

\def\beq{\begin{eqnarray}}
\def\eeq{\end{eqnarray}}

\newcommand{\x}{{\bf x}}
\newcommand{\y}{{\bf y}}
\newcommand{\z}{{\bf z}}

\begin{document}
\title{Asymptotic Expansion of the Heat Kernel Trace of Laplacians with Polynomial Potentials}

\author{Guglielmo Fucci\footnote{Electronic address: fuccig@ecu.edu}}

\affiliation{Department of Mathematics, East Carolina University, Greenville, NC 27858 USA}

\date{\today}
\vspace{2cm}
\begin{abstract}

It is well-known that the asymptotic expansion of the trace of the heat kernel for Laplace operators on smooth compact Riemmanian
manifolds can be obtained through termwise integration of the asymptotic expansion of the on-diagonal heat kernel. It is the purpose of this work to 
show that, in certain circumstances, termwise integration can be used to obtain the asymptotic expansion of the heat kernel trace 
for Laplace operators endowed with a suitable polynomial potential on unbounded domains. This is achieved by utilizing a resummed form of the asymptotic expansion 
of the on-diagonal heat kernel.
\end{abstract}


\maketitle

\section{Introduction}

The heat kernel is one of the most widely used spectral functions in many areas of mathematics and physics \cite{vassilevich03}.
For instance, in the ambit of spectral geometry, the heat kernel provides an invaluable tool 
for the analysis of the geometry of a Riemannian manifold \cite{gilkey95,gilkey04}. In addition, the heat kernel plays a pivotal 
role in the study of the one-loop effective action in quantum field theory and quantum gravity \cite{avramidi00,DW2003,kirsten01}.
The heat kernel is usually introduced as follows: Let $\mathscr{M}$ be a smooth compact Riemannian manifold, and let $\mathcal{V}$ be an Hermitian 
vector bundle over $\mathscr{M}$. We denote by $\mathcal{P}$ a Laplace-type operator acting on the space of smooth sections of $\mathcal{V}$, namely $C^{\infty}(\mathcal{V})$ . 
Under these assumptions the operator $\mathcal{P}$ is essentially selfadjoint \cite{gilkey95}. If $\partial M\neq \emptyset$ one considers 
boundary conditions that ensure the existence of a unique selfadjoint extension for $\mathcal{P}$. 
For $t>0$ the one-parameter 
family of operators 
\begin{equation}\label{00}
K(t)=\exp\left(-t\mathcal{P}\right)\;,
\end{equation}
forms a semigroup of bounded operators on $L^{2}(\mathcal{V})$, the space of square integrable sections of $\mathcal{V}$, called the heat semigroup. 
Let $\varphi_{n}\in L^{2}(\mathcal{V})$, with $n\in\mathbb{N}^{+}$, be the eigenfunctions of $\mathcal{P}$ and let $\lambda_{n}\in\mathbb{R}$ 
be the corresponding eigenvalues which form an increasing sequence bounded from below. By using the above notation, one defines the heat 
kernel $K(t|x,x')\in C^{\infty}((0,\infty)\times M\times M)$ associated with the heat semigroup (\ref{00}) as \cite{gilkey95}
\begin{equation}\label{01}
K(t|x,x^\prime)=\sum_{n=1}^{\infty}e^{-t\lambda_{n}}\varphi_{n}(x)\otimes \varphi^{\ast}_{n}(x^\prime)\;,
\end{equation} 
where $\lambda_{n}$ are counted with their algebraic multiplicity. The function (\ref{01}) satisfies the following parabolic partial differential equation
\begin{equation}
\left(\partial_{t}+\mathcal{P}\right)K(t|x,x^\prime)=0\;,
\end{equation} 
with the initial condition
\begin{equation}
K(0|x,x^\prime)=\delta(x,x^\prime)\;,
\end{equation}
where $\delta(x,x^\prime)$ denotes the covariant delta function, and appropriate boundary conditions if $\partial\mathscr{M}\neq\emptyset$. The semigroup (\ref{00}) can be 
proved to belong to the set of trace-class operators. This implies that the trace of (\ref{00}) exists and is defined as
\begin{equation}\label{0}
\textrm{Tr}_{L^{2}}e^{-t\mathcal{P}}=\int_{\mathscr{M}}\textrm{Tr}_{\mathcal{V}}K(t|x,x)\diff\textrm{vol}\;,
\end{equation}
where $\textrm{Tr}_{\mathcal{V}}$ represents the vector bundle trace, and $\diff\textrm{vol}$ is the volume element of the Riemannian manifold. 

The vast majority of research that focuses on the heat kernel has been performed for operators $\mathcal{P}$ on smooth compact
Riemannian manifolds with or without boundary (see e.g. \cite{byt,gilkey04,kirsten01,vassilevich03} and references therein). 
Complete results are also available for the coefficients of the heat kernel expansion of a Laplace-type operator endowed with an integrable potential 
defined on an unbounded domain. In this case the computation is performed by utilizing scattering theory and the 
appropriate Jost functions or, equivalently, the phase shift \cite{muller98,kirsten01}. 
Of particular importance, especially in the ambit of quantum mechanics and quantum field theory, is the Schr\"{o}dinger operator and its associated semigroup. 
A Schr\"{o}dinger operator, which we will denote by ${\cal H}$, is a Laplace type operator having the following general form \cite{berezin}
\begin{equation}\label{1}
{\cal H}=-\Delta+V(\x)\;,
\end{equation}
where $\Delta=\sum_{i=1}^{n}\partial^{2}/\partial x_{i}^{2}$ and the real function $V(\x)$ describes a potential. In many cases of physical interest
the Schr\"{o}dinger operator (\ref{1}) acts on suitable functions defined on the entire space $\mathbb{R}^{n}$. Although ${\cal H}$ is a Laplace type operator 
all the properties outlined earlier in this section, which are enjoyed by the operator ${\cal P}$ and the semigroup $e^{-t{\cal P}}$, do not 
immediately and trivially transfer to the Schr\"{o}dinger operator ${\cal H}$ and its associated semigroup $K(t)=e^{-t{\cal H}}$. This is mainly due to the fact that 
the operator ${\cal H}$ is defined on an unbounded domain. It can be proved, however, that for a wide variety of potentials of physical 
interest the Schr\"{o}dinger semigroup $e^{-t{\cal H}}$ is not only well defined but also a trace-class operator for all $t>0$.  

In fact, if $V(\x)\in L^{\infty}_{\textrm{loc}}(\mathbb{R}^{n})$ 
is a measurable locally bounded real-valued and positive function in $\mathbb{R}^{n}$ for which $\lim_{|x|\to\infty}V(\x)=+\infty$ 
and $V_{0}=\inf_{\x\in\mathbb{R}^{n}}V(\x)\geq0$ then the operator ${\cal H}$ is essentially selfadjoint, and its spectrum is real, 
discrete, with finite multiplicity and forms 
an increasing sequence approaching infinity \cite{berezin,alziary09,reed}. 
In addition, the eigenfunctions of ${\cal H}$ form an orthogonal basis of the space $L^{2}(\mathbb{R}^{n})$. 

Under these assumptions, the operator ${\cal H}$ generates, when $t>0$, a strongly continuous (heat)
semigroup $K(t)$ \cite{lax} for which the associated integral (heat) kernel 
$K(t|\x,\x')\in C^{\infty}((0,\infty)\times\mathbb{R}^{n}\times\mathbb{R}^{n})$ can be found through a 
representation in terms of the Feynman-Kac formula \cite{simon82,simon05}. In this work we will be mainly concerned with the 
analysis of the small-$t$ asymptotic expansion of the trace of the heat kernel for Schr\"{o}dinger operators ${\cal H}$. For this reason,
we will restrict the set of allowable potentials $V(\x)$ to those for which the heat semigroup is of trace-class.
In \cite{alziary09} one can find sufficient conditions that need to be imposed on the potential $V(\x)$ in order for the semigroup $K(t)$ to 
be intrinsically ultracontractive; a property which implies that $K(t)$ is a trace class operator (see e.g. \cite{davies84} Appendix A).
In this work we will consider smooth radially symmetric polynomial potentials, namely $V(|\x|)\in\mathcal C^{\infty}(\mathbb{R}^{n})$ such that 
$V(\x)=V(|\x|)$ and, in addition, $\lim_{|x|\to\infty}V(\x)=+\infty$ and $V_{0}=\inf_{\x\in\mathbb{R}^{n}}V(\x)\geq0$. For radially symmetric smooth polynomial
potentials for which $V_{0}\geq0$ and satisfy the abovementioned properties of growth at infinity, the semigroup $K(t)$ is intrinsically ultracontractive, except for the harmonic oscillator potential for which the semigroup is 
intrinsically hypercontractive \cite{alziary09,carmona79,davies84,davies90}. In both cases, however, it can be proved that 
$K(t)$ is of trace class, that $K(t|\x,\x')\in L^{2}((0,\infty)\times\mathbb{R}^{n}\times\mathbb{R}^{n})$ \cite{davies84}, and that the trace is given by the formula \cite{berezin}
\begin{equation}\label{2}
\textrm{Tr}_{L^{2}}e^{-t{\cal H}}=\int_{\mathbb{R}^{n}}K(t|\x,\x)\diff\x\;.
\end{equation}
The outline of the paper is as follows. In the next section we construct a suitable parametrix for the heat kernel of the
Schr\"{o}dinger operator ${\cal H}$. In Section \ref{sec3} we then prove that the obtained parametrix is indeed the small-$t$ asymptotic expansion 
of the Schr\"{o}dinger heat kernel. In Section \ref{sec4} we prove that termwise integration of the small-$t$ asymptotic expansion 
of the on-diagonal Schr\"{o}dinger heat kernel provides the small-$t$ asymptotic expansion of its trace. We then illustrate the termwise integration method
by computing the asymptotic expansion of the trace of the heat kernel associated with the one-dimensional Schr\"{o}dinger operator 
containing specific spherically symmetric polynomial potentials. The Conclusions summarize the main results and point to a few directions for future research.

\section{The Schr\"{o}dinger heat kernel and its parametrix} \label{Section2}

For Laplace type operators ${\cal P}$ defined on a smooth compact Riemannian manifold $M$ of dimension $d$ with or without boundary
it can be shown, in particular, that there exists a small-$t$ asymptotic expansion of the associated on-diagonal heat kernel \cite{gilkey95,mina53,mina49,seel68-10-288,seel66,seel69} of the form
\begin{equation}\label{3}
K(t|,x,x)\sim\frac{1}{(4\pi t)^{d/2}}\sum_{k=0}^{\infty}t^{k}b_{k}(x)\;,
\end{equation} 
where $b_{k}(x)$ are universal coefficients constructed from local geometric invariants of the manifold $M$ and its boundary $\partial M$.
The small-$t$ asymptotic expansion of the trace of the heat kernel is then obtained by integrating each coefficient $b_{k}(x)$ over the 
manifold $M$ and over its boundary $\partial M$. In more details one has 
\begin{equation}\label{4}
K(t)\sim \frac{1}{(4\pi t)^{d/2}}\sum_{k=0}^{\infty}t^{k}b_{k}\;,
\end{equation}
where $b_{k}$ is written in terms of the sum of a volume and boundary integral of the local invariants \cite{branson90,gilkey83,gilkey95}
\begin{equation}\label{5}
b_{k}=\int_{M} b_{k}(x) \diff x+\sum_{m=0}^{2k-1}\int_{\partial M} N^{m} b_{k,m}(y) \diff y\;,
\end{equation}
where $N^{m}$ is the $m$-th normal covariant derivative \cite{kirsten01}. 

At this point we consider the case of a Schr\"{o}dinger operator ${\cal H}$ defined on $\mathbb{R}^{n}$. 
The operator ${\cal H}$ is an elliptic second-order partial differential operator for which it is assumed that the potential $V(\x)\in L^{\infty}_{\textrm{loc}}(\mathbb{R}^{n})$ 
is smooth and satisfies the conditions $\lim_{|x|\to\infty}V(\x)=+\infty$ and $V_{0}=\inf_{\x\in\mathbb{R}^{n}}V(\x)\geq0$. These represent a set of sufficient conditions which allow one to shown 
that the small-$t$ asymptotic expansion of the on-diagonal heat kernel associated with the operator ${\cal H}$
has a form similar to the one displayed in (\ref{3}) \cite{bolte13,hitrik03}, namely
\begin{equation}\label{6}
K(t|\x,\x)\sim\frac{1}{(4\pi t)^{n/2}}\sum_{k=0}^{\infty}t^{k}a_{k}(\x)\;.
\end{equation}
If the heat semigroup generated by ${\cal H}$ has also the additional property of being of trace class, then one can analyze 
the small-$t$ asymptotic expansion of the corresponding trace.  
The small-$t$ asymptotic expansion of the trace of the 
heat semigroup $K(t)$ \emph{cannot}, however, be obtained by simply integrating over $\mathbb{R}^{n}$ the local coefficients $a_{k}(\x)$ of 
the expansion of the on-diagonal heat kernel $K(t|\x,\x)$. In fact, the local coefficients $a_{k}(\x)$ are constructed with invariants containing 
powers of the potential function $V(|\x|)$ and its derivatives \cite{cognola06,gilkey95,gilkey04,kirsten01,vassilevich03}. Integrating these coefficients over $\mathbb{R}^{n}$
would lead to meaningless divergent quantities since the potential $V(|\x|)$ is assumed to satisfy the condition 
$\lim_{|x|\to\infty}V(|\x|)=+\infty$ and is, therefore, non-integrable. 
The main purpose of this work is to propose 
a method that overcomes the abovementioned difficulties and that allows for the derivation of the small-$t$ 
asymptotic expansion of the trace of $K(t)$ from the one for the on-diagonal heat kernel $K(t|\x,\x)$ 
for the case of Schr\"{o}dinger operators on unbounded domains endowed with polynomial potentials satisfying the conditions stated in the previous section. 
Under these assumptions the semigroup is trace class and its trace is given by the expression in (\ref{2}).    

As a starting point we consider, instead of the asymptotic expansion (\ref{6}), its partially resummed form \cite{jack85,parker85}
\begin{equation}\label{7}
K(t|\x,\x)\sim\frac{1}{(4\pi t)^{\frac{n}{2}}}e^{-tV(|\x|)}\sum_{k=0}^{\infty}t^{k}A_{k}(|\x|)\;,
\end{equation}
where all the powers of $V(|\x|)$ in the expansion have been ``summed" into the exponential factor and the coefficients $A_{k}(|\x|)$ contain
only the derivatives of the potential $V(|\x|)$. 
If we were studying the trace of the heat kernel associated with the Laplacian on a smooth compact Riemannian manifold 
with no boundary, then its small-$t$ asymptotic expansion could be obtained by simply performing a termwise integration of the 
small-$t$ asymptotic expansion of the on-diagonal heat kernel \cite{gilkey95,gilkey04,kirsten01}. It is tempting, then, to
utilize a similar argument to obtain the small-$t$ asymptotic expansion of $K(t)$ from the one of $K(t|\x,\x)$ in (\ref{7}) since
the term by term integration of (\ref{7}) can be formally performed. In fact, 
the presence of the negative exponent in (\ref{7}) guarantees the convergence of the integrals when $t>0$ because $V(|\x|)\to+\infty$ 
for $|\x|\to+\infty$. This procedure, however, seems to be not justified at first since
in order for the small-$t$ asymptotic expansion of $K(t)$ to be given by the termwise integration of the expansion of $K(t|\x,\x)$ in (\ref{7})
not only all the ensuing integrals must be finite but also the expansion (\ref{7}) needs to be \emph{uniform} with respect to the integration variable $\x$ 
(see e.g. \cite{bleistein}, Theorem 1.7.5). Although the latter condition is a sufficient one for termwise 
integration, it is not necessary. Indeed it can be proved that uniformity 
of the asymptotic expansion is a condition that is too strong and can be relaxed \cite{lopez}.  
 
In what follows we will show that even though the remainder $R_{N}(t,\x)$ of the asymptotic expansion (\ref{7}) depends explicitly on the 
variable $\x$, namely
\begin{equation}\label{8}
K(t|\x,\x)=\frac{1}{(4\pi t)^{\frac{n}{2}}}e^{-tV(|\x|)}\sum_{k=0}^{N}t^{k}A_{k}(|\x|)+R_{N}(t,\x)\;,
\end{equation}
its integral satisfies the property
\begin{equation}\label{9}
\int_{\mathbb{R}^{n}}R_{N}(t,\x)\diff\x=O(t^{N-\frac{n}{2}+\alpha})\;,
\end{equation}     
with $\alpha>0$, and, hence, the asymptotic expansion of $K(t)$ can be obtained from the expression
\begin{equation}\label{10}
K(t)=\frac{1}{(4\pi t)^{\frac{n}{2}}}\sum_{k=0}^{N}t^{k}\left(\int_{\mathbb{R}^{n}}e^{-tV(|\x|)}A_{k}(|\x|)\diff\x\right)+O(t^{N-\frac{n}{2}+\alpha})\;,
\end{equation}
by performing the Laurent expansion of the integral in parentheses around $t=0$.   

The first step of our analysis consists in the explicit evaluation of the remainder $R_{N}(t,\x)$ of the asymptotic expansion in (\ref{7}). 
This can be achieved by constructing a parametrix $k_{N}(t|\x,\y)$ of order $N$ for the following heat equation \cite{bolte13}
\begin{equation}\label{11}
\left(\frac{\partial}{\partial t}+{\cal H}\right)K(t|\x,\y)=0\;\quad\textrm{with}\quad \lim_{t\to 0}K(t|\x,\y)=\delta(\x,\y)\;.
\end{equation}
Since the parametrix $k_{N}(t|\x,\y)$ is essentially a solution of (\ref{11}) valid for small values of $t$, we consider 
an ansatz of the form
\begin{equation}\label{12}
k_{N}(t|\x,\y)=\frac{1}{(4\pi t)^{\frac{n}{2}}}e^{-\frac{1}{t}p(\x,\y)-tV(|\x|)}\sum_{k=0}^{N+1}t^{k}A_{k}(\x,\y)\;.
\end{equation} 
By utilizing the above ansatz in the equation (\ref{11}) and by recalling that ${\cal H}=-\Delta+V(|\x|)$ we obtain
the relation
\begin{equation}\label{13}
\frac{e^{-\frac{1}{t}p}}{(4\pi t)^{\frac{n}{2}}}\left[\frac{1}{t^{2}}\left(p-(\nabla p)^{2}\right)+\frac{1}{t}\left(\Delta p+2(\nabla p)\cdot\nabla-\frac{n}{2}\right)
+\frac{\partial}{\partial t}-\Delta+V\right]\left(e^{-tV}\sum_{k=0}^{N+1}t^{k}A_{k}(\x,\y)\right)=0
\end{equation}  
where, for typographical convenience, we have set $p=p(\x,\y)$ and $V=V(|\x|)$. The equation (\ref{13}) can be solved recursively for $p(\x,\y)$ and $A_{k}(\x,\y)$. 
First, we choose the function $p(\x,\y)$ so that the most singular term as $t\to 0$ in (\ref{13}) vanishes. This leads to the Hamilton-Jacobi equation 
\begin{equation}\label{14}
p-(\nabla p)^{2}=0\;,
\end{equation}
whose solution is \cite{DW2003} 
\begin{equation}\label{15}
p(\x,\y)=\frac{1}{4}(\x-\y)^{2}\;.
\end{equation} 
By using the expression (\ref{15}) for $p(\x,\y)$ and by explicitly computing the action of the operator in (\ref{13}) we obtain the 
relation
\begin{eqnarray}\label{16}
\frac{e^{-\frac{1}{4t}(\x-\y)^{2}-tV}}{(4\pi t)^{\frac{n}{2}}}\sum_{k=0}^{N+1}\Big\{t^{k-1}[(\x-\y)\cdot\nabla+k]-t^{k}[(\x-\y)\cdot\nabla V+\Delta]\nonumber\\
+t^{k+1}[\Delta V+2(\nabla V)\cdot\nabla]-t^{k+2}(\nabla V)^{2}\Big\}A_{k}(\x,\y)=0\;.
\end{eqnarray}
By setting equal to zero the coefficient of each power of $t$ in equation (\ref{16}) we obtain a set of transport equations that allow us to compute recursively the 
functions $A_{k}(\x,\y)$. From the coefficient of $t^{-1}$ in (\ref{16}) we get
\begin{equation}\label{17}
(\x-\y)\cdot\nabla A_{0}(\x,\y)=0\;,
\end{equation} 
which is satisfied, by using the initial condition in (\ref{11}), when $A_{0}(\x,\y)=1$. 

From the coefficient of $t^{0}$ we obtain the transport equation
\begin{equation}\label{18}
(\x-\y)\cdot\nabla A_{1}(\x,\y)+A_{1}(\x,\y)-(\x-\y)\cdot\nabla V=0\;.
\end{equation}
where we have used the fact that $A_{0}(\x,\y)=1$. By setting $\x(s)=\y+s(\x-\y)$ with $s\in[0,1]$ \cite{bolte13}, one has $(\x-\y)\cdot\nabla=\diff/\diff s$
and equation (\ref{18}) becomes
\begin{equation}\label{19}
\left(\frac{\diff}{\diff s}+1\right)A_{1}(\x(s),\y)-\frac{\diff}{\diff s}V(|\x(s)|)=0\;.
\end{equation}
The solution of (\ref{19}) can be found in terms of an integral and has the form 
\begin{equation}\label{20}
A_{1}(\x,\y)=V(|\x|)-\int_{0}^{1}V(|\y+s(\x-\y)|)\diff s\;.
\end{equation}

The vanishing of the coefficient of $t$ in (\ref{16}) leads to the following equation determining $A_{2}(\x,\y)$ 
\begin{equation}\label{21}
\left[(\x-\y)\cdot\nabla+2\right]A_{2}(\x,\y)=\left[(\x-\y)\cdot\nabla V+\Delta-\Delta V\right]A_{1}(\x,\y)\;,
\end{equation} 
where, once again, we have used $A_{0}(\x,\y)=1$. By using the explicit expression for $A_{1}(\x,\y)$ in (\ref{20})
we can write the solution for $A_{2}(\x,\y)$ as
\begin{eqnarray}\label{22}
A_{2}(\x,\y)&=&\int_{0}^{1}s^{2}V(|\x(s)|)(\x-\y)\cdot\nabla V(|\x(s)|)\diff s\nonumber\\
&-&\int_{0}^{1}s^{2}(\x-\y)\cdot\nabla V(|\x(s)|)\left[\int_{0}^{1}V(|\y+s'(\x(s)-\y)|)\diff s'\right]\diff s\nonumber\\
&-&\int_{0}^{1}s\left[\int_{0}^{1}{s'}^{2}\Delta V(|\y+s'(\x(s)-\y)|)\diff s'\right]\diff s\;.
\end{eqnarray}

From the vanishing of the coefficients of $t^{k}$ with $2\leq k\leq N$ we obtain the relations
\begin{equation}\label{23}
\left[(\x-\y)\cdot\nabla+k+1\right]A_{k+1}(\x,\y)=q_{k}(\x,\y)\;,
\end{equation}
where we have defined the function
\begin{equation}\label{24}
q_{k}(\x,\y)=A_{k}(\x,\y)(\x-\y)\cdot\nabla V+\Delta A_{k}(\x,\y)-A_{k-1}(\x,\y)\Delta V-2(\nabla V)\cdot(\nabla A_{k-1}(\x,\y))+A_{k-2}(\x,\y)(\nabla V)^{2}\;.
\end{equation}

By using the expression for $k_{N}(t|\x,\y)$ in (\ref{12}) with $p(\x,\y)$ given by (\ref{15}) and the coefficients $A_{k}(\x,\y)$ found above
we get 
\begin{equation}\label{24a}
k_{N}(t|\x,\y)=\frac{1}{(4\pi t)^{\frac{n}{2}}}e^{-\frac{(\x-\y)^{2}}{4t}-tV(|\x|)}\sum_{k=0}^{N+1}t^{k}A_{k}(\x,\y)\;.
\end{equation}
which satisfies the relation
\begin{equation}\label{25}
\left(\frac{\partial}{\partial t}+{\cal H}\right)k_{N}(t|\x,\y)=R_{N}(t,\x,\y)\;,
\end{equation}
where $R_{N}(t,\x,\y)$ is defined as
\begin{equation}\label{26}
R_{N}(t,\x,\y)=-\frac{e^{-\frac{1}{4t}(\x-\y)^{2}-tV}}{(4\pi t)^{\frac{n}{2}}}q_{N+1}(\x,\y)t^{N+1}\;.
\end{equation}
When $N\geq n/2-1$ the function $k_{N}(t|\x,\y)\in C^{\infty}([0,\infty),\mathbb{R}^{n},\mathbb{R}^{n})$ is a parametrix for the kernel $K(t|\x,\y)$ \cite{bolte13}, 
which means, in particular, that $k_{N}(t|\x,\y)$ provides an approximation of $K(t|\x,\y)$ when $\x$ and $\y$ are close and when $t$ is small.

\section{Construction of the heat kernel asymptotic expansion}\label{sec3}

The parametrix $k_{N}(t|\x,\y)$ found in the previous section can be used to generate the small-$t$ asymptotic expansion of the 
Schr\"{o}dinger heat kernel $K(t|\x,\y)$. In addition, the function $R_{N}(t,\x,\y)$ in (\ref{26}) can be proved to yield an expression for
the remainder of the asymptotic expansion of $K(t|\x,\y)$. 

While the Schr\"{o}dinger heat kernel $K(t|\x,\y)$ satisfies the heat equation (\ref{11}), the parametrix $k_{N}(t|\x,\y)$
satisfies the associated non-homogeneous equation (\ref{25}) with the initial condition $\lim_{t\to 0}k_{N}(t|\x,\y)=\delta(\x,\y)$.
According to Duhamel's principle, the solution $k_{N}(t|\x,\y)$ of the non-homogeneous heat equation can be written in terms of the solution $K(t|\x,\y)$ of the associated
homogeneous one as follows \cite{taylor}
\begin{equation}\label{27}
k_{N}(t|\x,\y)=K(t|\x,\y)+\int_{0}^{t}\int_{\mathbb{R}^{n}}K(t-\tau|\x,\z)R_{N}(\tau,\z,\y)\diff\z\diff\tau\;.
\end{equation}
By utilizing the convolution of $f(t,\x,\y)$ and $g(t,\x,\y)$ \cite{bolte13}
\begin{equation}\label{28}
(f\ast g)(t,\x,\y)=\int_{0}^{t}\int_{\mathbb{R}^{n}}f(t-\tau,\x,\z)g(\tau,\z,\y)\diff\z\diff\tau\;,
\end{equation}
valid for any two functions for which the integral on the right hand side is well-defined, the solution (\ref{27}) can be rewritten as
\begin{equation}\label{29}
k_{N}(t|\x,\y)=K(t|\x,\y)+\left(K\ast R_{N}\right)(t,\x,\y)\;.
\end{equation}

At this point, the equation (\ref{29}) relating $k_{N}(t|\x,\y)$ and $K(t|\x,\y)$ can be formally solved for $K(t|\x,\y)$ to give
\begin{equation}\label{30}
K(t|\x,\y)=\left[k_{N}\ast\left(1+R_{N}\right)^{-1}\right](t|\x,\y)\;.
\end{equation}
In order to define the quantity $\left(1+R_{N}\right)^{-1}$ we use the geometric series and obtain the following formal expression for the 
Schr\"{o}dinger heat kernel \cite{bolte13}
\begin{equation}\label{31}
K(t|\x,\y)=k_{N}(t|\x,\y)+\left[k_{N}\ast\sum_{l=1}^{\infty}(-1)^{l}(R_{N})^{\ast l}\right](t|\x,\y)\;,
\end{equation}
where $(R_{N})^{\ast l}$ denotes the function $R_{N}(t,\x,\y)$ convoluted with itself $l$ times.
We are left, now, with the task of proving that this formal procedure indeed gives the desired asymptotic expansion 
of the Schr\"{o}dinger heat kernel. First, one needs to prove that the infinite series in (\ref{31}) does converge uniformly to a function 
$Q(t,\x,\y)\in  C^{\infty}([0,\infty),\mathbb{R}^{n},\mathbb{R}^{n})$. One, then, needs to show that the expression for $K(t|\x,\y)$
given in (\ref{31}) solves the initial value problem (\ref{11}) and, finally, that (\ref{31}) provides a small-$t$ 
asymptotic expansion of $K(t|\x,\y)$ \cite{rosenberg}. 
 
By utilizing the definition (\ref{28}), the term $(R_{N})^{\ast l}$ can be written explicitly as
\begin{eqnarray}\label{32}
(R_{N})^{\ast l}(t,\x,\y)&=&\int_{0}^{t}\int_{0}^{t_{1}}\cdots\int_{0}^{t_{l-2}}\int_{\mathbb{R}^{n}}\cdots\int_{\mathbb{R}^{n}}R_{N}(t-t_{1},\x,\z_{1})R_{N}(t_{1}-t_{2},\z_{1},\z_{2})\nonumber\\
&\times&R_{N}(t_{2}-t_{3},\z_{2},\z_{3})\cdots R_{N}(t_{l-1},\z_{l-1},\y)\diff\z_{l-1}\diff\z_{l-2}\cdots\diff\z_{1}\diff t_{l-1}\diff t_{l-2}\cdots\diff t_{1}\;,
\end{eqnarray} 
which leads to the inequality
\begin{eqnarray}\label{33}
\left|(R_{N})^{\ast l}(t,\x,\y)\right|&\leq& \int_{0}^{t}\int_{0}^{t_{1}}\cdots\int_{0}^{t_{l-2}}\int_{\mathbb{R}^{n}}\cdots\int_{\mathbb{R}^{n}}\left|R_{N}(t-t_{1},\x,\z_{1})\right|\left|R_{N}(t_{1}-t_{2},\z_{1},\z_{2})\right|\nonumber\\
&\times&\left|R_{N}(t_{2}-t_{3},\z_{2},\z_{3})\right|\cdots \left|R_{N}(t_{l-1},\z_{l-1},\y)\right|\diff\z_{l-1}\diff\z_{l-2}\cdots\diff\z_{1}\diff t_{l-1}\diff t_{l-2}\cdots\diff t_{1}\;.
\end{eqnarray}
Now, from (\ref{26}) we have the estimate
\begin{equation}\label{34}
\left|R_{N}(t,\x,\y)\right|\leq t^{N-\frac{n}{2}+1}e^{-\frac{1}{4t}(\x-\y)^{2}-tV}\left|q_{N+1}(\x,\y)\right|\;.
\end{equation}
Since the functions $q_{k}(\x,\y)$ are constructed from powers of the potential $V$ and its derivatives (cft. (\ref{24})) 
and since $V$ is a polynomial function satisfying the conditions stated at the beginning of the paper, we can conclude that $|q_{k}(\x,\y)|$ is a polynomial function as well. 
When $\x\neq\y$, which is the only case we need to focus on for the purpose of estimating the integrals in the convolution, 
one can utilize Lemma (4.1) of \cite{bolte13} to conclude that for all $\epsilon>0$ and small enough there exist $\alpha=1-\epsilon$
and a constant $C_{N}$ such that (\ref{34}) can be written as 
\begin{equation}\label{35}
\left|R_{N}(t,\x,\y)\right|\leq C_{N}t^{N-\frac{n}{2}+1}e^{-\frac{\alpha}{4t}(\x-\y)^{2}- tV}\;.
\end{equation} 
By using the last estimate in the inequality (\ref{33}) and the fact that for $n\geq 2$, $\exp\{-(t_{n-2}-t_{n-1})V\}\leq 1$, we obtain
\begin{eqnarray}\label{36}
\left|(R_{N})^{\ast l}(t,\x,\y)\right|&\leq&M_{N}\int_{0}^{t}\int_{0}^{t_{1}}\cdots\int_{0}^{t_{l-2}}\int_{\mathbb{R}^{n}}\cdots\int_{\mathbb{R}^{n}}
(t-t_{1})^{N+1-\frac{n}{2}}(t_{1}-t_{2})^{N+1-\frac{n}{2}}\cdots t_{l-1}^{N+1-\frac{n}{2}}e^{-(t-t_{1})V}\nonumber\\
&\times&e^{-\frac{\alpha}{4(t-t_{1})}(\x-\z_{1})^{2}}e^{-\frac{\alpha}{4(t_{1}-t_{2})}(\z_{1}-\z_{2})^{2}}\cdots e^{-\frac{\alpha}{4t_{l-1}}(\z_{l-1}-\y)^{2}}\diff\z_{l-1}\diff\z_{l-2}\cdots\diff\z_{1}\diff t_{l-1}\diff t_{l-2}\cdots\diff t_{1}\;.\;\;\;\;
\end{eqnarray} 
The integrals over the variables $\z_{i}$ represent simply a convolution of gaussian functions and can be computed by iteration by noticing that
for all $a>0$ and $b>0$ one has
\begin{equation}\label{37}
\int_{\mathbb{R}^{n}}e^{-a(\x-\z)^{2}}e^{-b(\z-\y)^{2}}\diff\z=\left(\frac{\pi}{a+b}\right)^{\frac{n}{2}}e^{\frac{ab}{a+b}(\x-\y)^{2}}\;.
\end{equation}  
The last remark allows us to write
\begin{eqnarray}\label{38}
\int_{\mathbb{R}^{n}}\cdots\int_{\mathbb{R}^{n}}e^{-\frac{\alpha}{4(t-t_{1})}(\x-\z_{1})^{2}}e^{-\frac{\alpha}{4(t_{1}-t_{2})}(\z_{1}-\z_{2})^{2}}\cdots e^{-\frac{\alpha}{4t_{l-1}}(\z_{l-1}-\y)^{2}}\diff\z_{l-1}\diff\z_{l-2}\cdots\diff\z_{1}\nonumber\\
=\left(\frac{2\pi}{\alpha}\right)^{\frac{n(l-1)}{2}}t^{-\frac{n}{2}}(t-t_{1})^{\frac{n}{2}}(t_{1}-t_{2})^{\frac{n}{2}}\cdots(t_{l-2}-t_{l-1})^{\frac{n}{2}}t_{l-1}^{\frac{n}{2}}
e^{-\frac{\alpha}{4t}(\x-\y)^{2}}\;.
\end{eqnarray}
The result obtained for the convolution of gaussian functions can be used in (\ref{36}) to get
\begin{eqnarray}\label{39}
\lefteqn{\left|(R_{N})^{\ast l}(t,\x,\y)\right|\leq M_{N}\left(\frac{2\pi}{\alpha}\right)^{\frac{n(l-1)}{2}}e^{-\frac{\alpha}{4t}(\x-\y)^{2}}t^{-\frac{n}{2}}}\nonumber\\
&&\times\int_{0}^{t}\int_{0}^{t_{1}}\cdots\int_{0}^{t_{l-2}}
(t-t_{1})^{N+1}(t_{1}-t_{2})^{N+1}\cdots t_{l-1}^{N+1}e^{-(t-t_{1})V}\diff t_{l-1}\diff t_{l-2}\cdots\diff t_{1}\;.
\end{eqnarray}
We are now left with the task of analyzing the integrals over the variables $t_{i}$. By performing the change of variables $t_{i}=ts_{i}$
we have
\begin{eqnarray}\label{40}
\lefteqn{\int_{0}^{t}\int_{0}^{t_{1}}\cdots\int_{0}^{t_{l-2}}
(t-t_{1})^{N+1}(t_{1}-t_{2})^{N+1}\cdots t_{l-1}^{N+1}e^{-(t-t_{1})V}\diff t_{l-1}\diff t_{l-2}\cdots\diff t_{1}}\nonumber\\
&&=t^{l(N+2)-1}\int_{0}^{1}\int_{0}^{s_{1}}\cdots\int_{0}^{s_{l-2}}
(1-s_{1})^{N+1}(s_{1}-s_{2})^{N+1}\cdots s_{l-1}^{N+1}e^{- t(1-s_{1})V}\diff s_{l-1}\diff s_{l-2}\cdots\diff s_{1}\;.
\end{eqnarray}  
Since $0<s_{l-1}<s_{l-2}<\cdots<s_{1}<1$ we can provide the following bound for the above integrals 
\begin{eqnarray}\label{41}
t^{l(N+2)-1}\int_{0}^{1}\int_{0}^{s_{1}}\cdots\int_{0}^{s_{l-2}}
(1-s_{1})^{N+1}(s_{1}-s_{2})^{N+1}\cdots s_{l-1}^{N+1}e^{- t(1-s_{1})V}\diff s_{l-1}\diff s_{l-2}\cdots\diff s_{1}\nonumber\\
\leq t^{l(N+2)-1}\int_{0}^{1}\int_{0}^{s_{1}}\cdots\int_{0}^{s_{l-2}}e^{-t(1-s_{1})V}\diff s_{l-1}\diff s_{l-2}\cdots\diff s_{1}\nonumber\\
=\frac{t^{l(N+2)-1}}{(l-2)!}\int_{0}^{1}e^{- t(1-s_{1})V}s_{1}^{l-2}\diff s_{1}\;.
\end{eqnarray}
By exploiting the second mean value theorem one can show that there exists $\gamma\in(0,1)$ such that 
\begin{equation}\label{42}
\frac{t^{l(N+2)-1}}{(l-2)!}\int_{0}^{1}e^{- t(1-s_{1})V}s_{1}^{l-2}\diff s_{1}=\frac{t^{l(N+2)-1}}{(l-1)!}e^{-tV}\gamma^{l-1}\;.
\end{equation}

Based on the last remarks we have the following inequality 
\begin{equation}\label{43}
\left|(R_{N})^{\ast l}(t,\x,\y)\right|\leq M_{N}\left(\frac{2\pi}{\alpha}\right)^{\frac{n(l-1)}{2}} \frac{e^{-tV}}{(l-1)!}e^{-\frac{\alpha}{4t}(\x-\y)^{2}}t^{-\frac{n}{2}+l(N+2)-1}\;,
\end{equation}
which is bounded as $t\to 0$ when $l\geq(n+2)/(2N+4)$. By using the estimate (\ref{43}) one can conclude that the
series 
\begin{equation}\label{44}
Q(t,\x,\y)=\sum_{l=1}^{\infty}(-1)^{l}(R_{N})^{\ast l}(t,\x,\y)\;,
\end{equation}
converges uniformly in $\x$, $\y$, and $t\in(0,\infty)$ when $l\geq(n+2)/(2N+4)$ to define a function $Q(t,\x,\y)\in  C^{\infty}([0,\infty),\mathbb{R}^{n},\mathbb{R}^{n})$.
This condition is satisfied for $l\in\mathbb{N}^{+}$ by simply considering $N\geq n/2-1$ terms in the expansion (\ref{24a}). 

From (\ref{44}) we can therefore conclude that the expression for the Schr\"{o}dinger heat kernel $K(t|\x,\y)$ in (\ref{31}) is well defined. We only need to show, now, 
that $K(t|\x,\y)$ in (\ref{31}) is indeed a solution of the initial value problem (\ref{11}). By using (\ref{31}), with the definition (\ref{44}), in the 
heat equation (\ref{11}) we obtain
\begin{eqnarray}\label{45}
\left(\frac{\partial}{\partial t}+{\cal H}\right)K(t|\x,\y)&=&\left(\frac{\partial}{\partial t}+{\cal H}\right)\left[k_{N}(t|\x,\y)+(k_{N}\ast Q)(t,\x,\y)\right]\nonumber\\
&=&R_{N}(t,\x,\y)+\left(\frac{\partial}{\partial t}+{\cal H}\right)(k_{N}\ast Q)(t,\x,\y)\;.
\end{eqnarray}
The last term of the chain of equalities in (\ref{45}) can be explicitly computed by recalling the definition of convolution in (\ref{28})
\begin{eqnarray}\label{46}
\left(\frac{\partial}{\partial t}+{\cal H}\right)\left[(k_{N}\ast Q)(t,\x,\y)\right]&=&\left(\frac{\partial}{\partial t}+{\cal H}\right)
\int_{0}^{t}\int_{\mathbb{R}^{n}}k_{N}(t-\tau|\x,\z)Q(\tau,\z,\y)\diff\z\diff\tau\nonumber\\
&=&Q(t,\x,\y)+\int_{0}^{t}\int_{\mathbb{R}^{n}}R_{N}(t-\tau|\x,\z)Q(\tau,\z,\y)\diff\z\diff\tau\nonumber\\
&=&Q(t,\x,\y)+(R_{N}\ast Q)(t,\x,\y)\;.
\end{eqnarray}
This result, once substituted in (\ref{45}), leads to the relation
\begin{equation}\label{47}
\left(\frac{\partial}{\partial t}+{\cal H}\right)K(t|\x,\y)=R_{N}(t,\x,\y)+Q(t,\x,\y)+(R_{N}\ast Q)(t,\x,\y)\;.
\end{equation}
The right-hand-side of the last equation can actually be simplified further. In fact, by recalling the definition of $Q(t,\x,\y)$ in (\ref{44}) we have that 
\begin{equation}\label{48}
 \left(\frac{\partial}{\partial t}+{\cal H}\right)K(t|\x,\y)=R_{N}(t,\x,\y)+\sum_{l=1}^{\infty}(-1)^{l}(R_{N})^{\ast l}(t,\x,\y)
 +\sum_{l=1}^{\infty}(-1)^{l}(R_{N})^{\ast (l+1)}(t,\x,\y)=0.
\end{equation}   
In addition, it is not difficult to prove that 
\begin{equation}\label{49}
\lim_{t\to 0}K(t|\x,\y)=\lim_{t\to 0}k_{N}(t|\x,\y)=\delta(\x,\y)\;.
\end{equation}
Equation (\ref{48}) and the limit (\ref{49}) hence show that the Schr\"{o}dinger heat kernel $K(t|\x,\y)$, given by the formula (\ref{31}),
is the solution to the initial value problem (\ref{11}).

In terms of the explicit expression for $k_{N}(t|\x,\y)$ given in (\ref{24a}) the Schr\"{o}dinger heat kernel $K(t|\x,\y)$ can be rewritten as 
\begin{equation}\label{50}
K(t|\x,\y)=\frac{1}{(4\pi t)^{\frac{n}{2}}}e^{-\frac{(\x-\y)^{2}}{4t}-tV(|\x|)}\sum_{k=0}^{N+1}t^{k}A_{k}(\x,\y)+r_{N}(t,\x,\y)\;,
\end{equation}
where we have introduced the notation
\begin{equation}\label{51}
r_{N}(t,\x,\y)=(k_{N}\ast Q)(t,\x,\y)\;.
\end{equation}

It is important, at this point, to analyze in more details the term $r_{N}(t,\x,\y)$. It is clear, from the definition (\ref{44})
of $Q(t,\x,\y)$, that the smallest power of $t$ in $r_{N}(t,\x,\y)$ comes from the convolution of $k_{N}(t|\x,\y)$ and $R_{N}(t,\x,\y)$.
To estimate this convolution we follow the same argument used to find a bound for $R_{N}(t,\x,\y)$. From the definition (\ref{28}) 
we have
\begin{eqnarray}\label{52}
\left|(k_{N}\ast R_{N})(t,\x,\y)\right|\leq \int_{0}^{t}\int_{\mathbb{R}^{n}}\left|k_{N}(t-\tau|\x,\z)\right|\left|R_{N}(\tau,\z,\y)\right|\diff\z\diff\tau\;.
\end{eqnarray}  
Moreover, according to (\ref{24a}) we can write, for small $t$,
\begin{equation}\label{52a}
\left|k_{N}(t|\x,\y)\right|\leq t^{-\frac{n}{2}}e^{-t V}e^{-\frac{1}{4t}(\x-\y)^{2}}\sum_{n=0}^{N+1}\left|A_{k}(\x,\y)\right|\;.
\end{equation}
The coefficients $A_{k}(\x,\y)$ are polynomial functions since they are constructed from the derivatives of $V$ and their powers. By using an argument similar to the 
one employed to obtain (\ref{35}) from (\ref{34}), one can prove that for all $\epsilon>0$ and small enough there exist $\alpha=1-\epsilon$
and a constant $c_{N}$ such that (\ref{52a}) can be expressed as
\begin{equation}\label{53}
\left|k_{N}(t|\x,\z)\right|\leq c_{N}t^{-n/2}e^{-t V}e^{-\frac{\alpha}{4t}(\x-\y)^{2}}\;.
\end{equation}
By exploiting the estimate in (\ref{35}) and the one in (\ref{53}) 
one can rewrite (\ref{52}) as
\begin{eqnarray}\label{54}
\left|k_{N}(t|\x,\y)\ast R_{N}(t,\x,\y)\right|&\leq& g_{N} \int_{0}^{t}\int_{\mathbb{R}^{n}} (t-\tau)^{-\frac{n}{2}}\tau^{-\frac{n}{2}+N+1}e^{-(t-\tau)V}
e^{-\frac{\alpha}{4(t-\tau)}(\x-\z)^{2}}e^{-\frac{\alpha}{4\tau}(\z-\y)^{2}}\diff\z\diff\tau\nonumber\\
&=&g_{N}\left(\frac{2\pi}{\alpha}\right)^{\frac{n}{2}}t^{-\frac{n}{2}}e^{-\frac{\alpha}{4t}(\x-\y)^{2}}\int_{0}^{t}\tau^{N+1}e^{-(t-\tau)V}\diff\tau\;,
\end{eqnarray}
where the equality is obtained by using (\ref{37}) and $g_{N}>0$ is a suitable constant. 
By changing variables $\tau=ts$ and by applying the second mean value theorem to the resulting integral 
one can show that there exists a constant $\delta_{N}>0$ such that 
\begin{equation}\label{55}
\left|(k_{N}\ast R_{N})(t,\x,\y)\right|\leq \delta_{N}t^{-\frac{n}{2}+N+2}e^{- tV}e^{-\frac{\alpha}{4t}(\x-\y)^{2}}\;. 
\end{equation}
The last bound proves that the smallest power of $t$ in $r_{N}(t,\x,\y)$ is $t^{-\frac{n}{2}+N+2}$ and, hence, the expression (\ref{50})
is a legitimate small-$t$ asymptotic expansion which can be written as 
\begin{equation}\label{56}
K(t|\x,\y)=\frac{1}{(4\pi t)^{\frac{n}{2}}}e^{-\frac{(\x-\y)^{2}}{4t}-tV(|\x|)}\sum_{k=0}^{N+1}t^{k}A_{k}(\x,\y)+O\left(t^{-\frac{n}{2}+N+2}\right)\;,
\end{equation}
where the remainder satisfies the non-uniform bound in the variables $\x$ and $\y$ displayed in (\ref{55}). 

\section{Asymptotic expansion of the trace of the Schr\"{o}dinger heat kernel}\label{sec4}

The small-$t$ asymptotic expansion of the on-diagonal Schr\"{o}dinger heat kernel can be easily obtained from the expression (\ref{50})
by performing the coincidence limit $\y\to\x$. It is then not very difficult to get the formula
\begin{equation}\label{57}
K(t|\x,\x)=\frac{1}{(4\pi t)^{\frac{n}{2}}}e^{-tV(|\x|)}\sum_{k=0}^{N+1}t^{k}A_{k}(|\x|)+r_{N}(t,\x)\;,
\end{equation}
where the remainder $r_{N}(t,\x)$ is bounded, according to (\ref{55}), as follows
\begin{equation}\label{58}
\left|r_{N}(t,\x)\right|\leq \delta_{N} t^{-\frac{n}{2}+N+2}e^{- tV(|\x|)}\;.
\end{equation}  
The trace of the Schr\"{o}dinger semigroup is given by the expression in (\ref{2}). By using the expansion (\ref{57}) we obtain
\begin{equation}\label{59}
K(t)=\frac{1}{(4\pi t)^{\frac{n}{2}}}\sum_{k=0}^{N+1}\left(\int_{\mathbb{R}^{n}}A_{k}(|\x|)e^{-tV(|\x|)}\diff\x\right)t^{k}+\int_{\mathbb{R}^{n}}r_{N}(t,\x)\diff\x\;.
\end{equation}

It is important at this point to analyze in detail the small-$t$ behavior of the integrals appearing in the sum in (\ref{59}).
Due to the fact that $V(|\x|)$ is radially symmetric we can use spherical coordinates and write 
$V(|\x|)=V(r)$ where $r>0$ denotes the radial coordinate. We define
\begin{equation}\label{60}
{\cal I}(k,t)=\int_{\mathbb{R}^{n}}A_{k}(|\x|)e^{-tV(|\x|)}\diff\x=S_{n}\int_{0}^{\infty} r^{n-1}A_{k}(r)e^{-tV(r)}\diff r\;,
\end{equation}
where $S_{n}=2\pi^{n/2}/\Gamma(n/2)$ is the result of the integration over the angular variables.
Since the potential $V(r)$ is assumed to be a
polynomial function one can write $V(r)$ in the form
\begin{equation}\label{60a}
V(r)=\sum_{j=0}^{q}c_{j}r^{j}\;.
\end{equation}
The potential has to be chosen such that the conditions stated at the beginning of the paper are satisfied. This implies that we need to assume that $c_{q}>0$, which guarantees that $V(r)\to\infty$ as $r\to\infty$, and that $c_{0}\geq 0$ is large enough so that $\textrm{inf}_{r>0}V(r)\geq 0$. 

Performing the change of coordinates $c_{q}tr^{q}=s$ in the integral in equation (\ref{60}) allows us
to express the integral ${\cal I}(k,t)$ in a form which is suitable for the analysis of its small-$t$ expansion 
\begin{equation}\label{61}
{\cal I}(k,t)=\frac{S_{n}}{q c_{q}^{n/q} t^{n/q}}\int_{0}^{\infty}s^{\frac{n}{q}-1}e^{-s}A_{k}\left(\left(\frac{s}{tc_{q}}\right)^{1/q}\right)
\exp\Bigg\{-t\sum_{j=0}^{q-1}c_{j}\left(\frac{s}{tc_{q}}\right)^{j/q}\Bigg\}\diff s\;.
\end{equation} 
The small-$t$ expansion, with $s$ fixed, of the exponential containing the polynomial in $(s/tc_{q})^{1/q}$can be computed from the relation
\begin{equation}\label{62}
\exp\Bigg\{-t\sum_{j=0}^{q-1}c_{j}\left(\frac{s}{tc_{q}}\right)^{j/q}\Bigg\}=\sum_{n=0}^{M}\frac{(-1)^{n}}{n!}\left(\sum_{j=0}^{q-1}c_{j}\left(\frac{s}{tc_{q}}\right)^{j/q}\right)^{n}t^{n}+O\left(t^{M+\frac{1}{q}}\right)\;.
\end{equation}
To rearrange the sum in (\ref{62}) we utilize the expression
\begin{equation}\label{62a}
\left(\sum_{j=0}^{q-1}c_{j}\left(\frac{s}{tc_{q}}\right)^{j/q}\right)^{n}=\sum_{l=0}^{(q-1)n}D_{l}^{n}\left(\frac{s}{tc_{q}}\right)^{\frac{l}{q}}\;,
\end{equation}
which can be obtained from the multinomial expansion (see e.g. \cite{abramo70,price46}) with coefficients
\begin{eqnarray}\label{62b}
D^{n}_{l}=\sum_{n_{0}=[l/q-1]}^{l}\sum_{n_{1}=0}^{l-n_{0}}\cdots\sum_{n_{q-3}=0}^{l-n_{0}-\ldots-n_{q-4}}\binom{n}{n_{0}}\binom{n_{0}}{n_{1}}\cdots\binom{n_{q-3}}{l-n_{0}-\ldots-n_{q-3}}\times\nonumber\\
c_{0}^{n-n_{0}}c_{1}^{n_{0}-n_{1}}\cdots c_{q-2}^{2n_{q-3}+n_{0}+\ldots+n_{q-4}-l}c_{q-1}^{l-n_{0}-n_{1}-\ldots-n_{q-3}}\;.
\end{eqnarray}
By utilizing (\ref{62a}) in the expression (\ref{62}) and by rearranging the ensuing sum in increasing powers of $t$ one obtains
\begin{equation}\label{62c}
\exp\Bigg\{-t\sum_{j=0}^{q-1}c_{j}\left(\frac{s}{tc_{q}}\right)^{j/q}\Bigg\}=\sum_{j=0}^{Mq}H_{j}(s)t^{\frac{j}{q}}+O\left(t^{M+\frac{1}{q}}\right)\;,
\end{equation}
where one needs to have $M\geq j$ in order to account for all the terms proportional to $t^{j/q}$. 
In (\ref{62c}), $H_{0}(s)=1$ and $H_{j}(s)$ with $j\geq 1$ are polynomials in $s^{1/q}$ of the form
\begin{equation}\label{62d}
H_{j}(s)=\sum_{p=\left\lceil \frac{j}{q}\right\rceil}^{j}\frac{(-1)^{p}}{p!}D_{pq-j}^{p}\left(\frac{s}{c_{q}}\right)^{\frac{pq-j}{q}}\;.
\end{equation}
In the last expression we have used the standard symbol $\lceil x\rceil$ to denote the ceiling function.

The next step consists in the study of the small-$t$ behavior of $A_{k}\left(\left(s/tc_{q}\right)^{1/q}\right)$. 
To accomplish this task we first need to derive some properties of the coefficients $A_{k}(r)$.
The general form of $A_{k}(r)$ can be identified by utilizing dimensional arguments \cite{kirsten01}. In this framework
the coefficients $A_{k}(r)$ have dimension $l^{2k}$ where $l$ represents a unit of length \cite{gilkey95,gilkey04,kirsten01}. 
By denoting by $p_{V}$ the powers of $V(r)$ and by $p_{\nabla}$ the powers of the derivative $\partial_{k}$
we have for each coefficient $A_{k}(r)$, $k\geq 0$, the relation \cite{kirsten01}
\begin{equation}\label{62e}
2p_{V}+p_{\nabla}=2k\;,
\end{equation}  
which holds true since the potential $V(r)$ and the derivative $\partial_{k}$ have dimension $l^{2}$ and $l$, respectively. 

At this point a few remarks are in order. As already mentioned earlier, $A_{k}(r)$ contains 
no powers of $V(r)$. This implies, in particular, that $p_{\nabla}>0$. In addition, since 
derivatives of higher powers of the potential, namely $V^{n}(r)$, can be written as a linear combination of the derivatives of $V(r)$, the 
independent invariants in $A_{k}(r)$ that can be constructed from the potential $V(r)$ must satisfy the constraint $p_{V}\leq p_{\nabla}$.
From the relation (\ref{62e}) it is not difficult to realize that the coefficients $A_{k}(r)$ contain an even number of derivatives and, therefore, have the general form in terms of the variable $(s/tc{q})^{1/q}$
\begin{equation}\label{63}
A_{k}\left(\left(\frac{s}{tc_{q}}\right)^{1/q}\right)=\sum_{l=0}^{\gamma_{k}}\Omega^{k}_{l}\left(\frac{s}{tc_{q}}\right)^{\frac{l}{q}}\;,
\end{equation}
where $\gamma_{k}$ is an integer which can be found by using (\ref{62a}). The terms $\Omega_{l}^{k}$ are real coefficients computable from the explicit expression 
of $A_{k}(r)$ which, in turn, are obtained from the recurrence relations for $A_{k}(\x,\y)$ given in Section \ref{Section2} by using $V(r)$ in (\ref{60a}).
The coefficient $\gamma_{k}$ provides the highest power of $r$ appearing in $A_{k}(r)$ which can be obtained by 
maximizing $p_{V}\in\mathbb{N}^+$ within the constraint $p_{V}\leq p_{\nabla}$ mentioned above. For $k=0$, and $k=1$ we use the fact that
$A_{0}(r)=1$ and that $A_{1}(r)=0$, obtained from (\ref{20}) once the limit $\y\to\x$ is performed, to conclude that $\gamma_{0}=\gamma_{1}=0$. For $k\geq 2$, the maximum value of  
$p_{V}$ is attained at $\bar{p}_{V}=\lfloor 2k/3\rfloor$ and the corresponding value of $p_{\nabla}\in\mathbb{N}^+$, satisfying the above mentioned constraint, is $\bar{p}_{\nabla}=2(k-\lfloor2k/3\rfloor)$, where $\lfloor x\rfloor$ denotes the floor function. From the above remarks we can conclude that for a potential $V(r)$ of the form 
displayed in (\ref{60a}) the highest power of $r$ entering $A_{k}(r)$ is $q\bar{p}_{V}-\bar{p}_{\nabla}$ which implies that 
\begin{equation}\label{64}
\gamma_{k}=\left\lfloor\frac{2k}{3}\right\rfloor(q+2)-2k\;,
\end{equation} 
for $k\geq 2$.

By using (\ref{62c}) and (\ref{63}) in the integral (\ref{61}) we obtain 
\begin{equation}\label{67}
{\cal I}(k,t)=\frac{S_{n}}{q c_{q}^{n/q} t^{\frac{n}{q}+\frac{\gamma_{k}}{q}}}\int_{0}^{\infty}s^{\frac{n}{q}-1}e^{-s}
\left(\sum_{l=0}^{\gamma_{k}}\Omega^{k}_{l}\left(\frac{s}{c_{q}}\right)^{\frac{l}{q}}t^{\frac{\gamma_{k}-l}{q}}\right)\left(\sum_{j=0}^{Mq}H_{j}(s)t^{\frac{j}{q}}\right)\diff s+O\left(t^{\frac{-n+Mq+1}{q}}\right)\;,
\end{equation}
The integral in (\ref{67}) can be explicitly computed once the product of the two sums has been performed and organized in increasing powers of $t$.
The Cauchy product formula allows us to rearrange the product of the sums in the previous equation as follows
\begin{equation}\label{67a}
\left(\sum_{l=0}^{\gamma_{k}}\Omega^{k}_{l}\left(\frac{s}{c_{q}}\right)^{\frac{l}{q}}t^{\frac{\gamma_{k}-l}{q}}\right)\left(\sum_{j=0}^{Mq}H_{j}(s)t^{\frac{j}{q}}\right)
=\sum_{p=0}^{Mq+\gamma_{k}}\Lambda_{p}^{k}(s)t^{\frac{p}{q}}\;,
\end{equation}
where the coefficients $\Lambda_{p}^{k}(s)$ can be found to have the form
\begin{equation}\label{67b}
\Lambda_{p}^{k}(s)=\sum_{n=\textrm{max}\{0,p-\gamma_{k}\}}^{p}\Omega^{k}_{\gamma_{k}-p+n}H_{n}(s)\left(\frac{s}{c_{q}}\right)^{\frac{\gamma_{k}-p+n}{q}}\;.
\end{equation}
By using the relation (\ref{67a}) in (\ref{67}) we obtain the desired small-$t$ asymptotic expansion
\begin{equation}\label{68}
{\cal I}(k,t)=\frac{S_{n}}{ c_{q}^{n/q} t^{\frac{n}{q}+\frac{\gamma_{k}}{q}}}\sum_{p=0}^{Mq+\gamma_{k}}T_{p}^{n,k}t^{\frac{p}{q}}+O\left(t^{\frac{-n+Mq+1}{q}}\right)\;,
\end{equation}
where the newly introduced coefficients $T_{p}^{n,k}$ have the form (cf. (\ref{67b}) and (\ref{62d})),
\begin{equation}\label{68a}
T_{p}^{n,k}=\frac{1}{q}\sum_{j=\textrm{max}\{0,p-\gamma_{k}\}}^{p}\sum_{l=\left\lceil \frac{j}{q}\right\rceil}^{j}
\frac{(-1)^{l}}{l!}\Omega_{\gamma_{k}-p+j}^{k}D_{lq-j}^{l}c_{q}^{\frac{p-\gamma_{k}-lq}{q}}\int_{0}^{\infty}s^{\frac{n+\gamma_{k}-p+lq}{q}-1}e^{-s}\diff s\;.
\end{equation}
By explicitly evaluating the elementary integral appearing in (\ref{68a}) we obtain 
\begin{equation}\label{68b}
T_{p}^{n,k}=\frac{1}{q}\sum_{j=\textrm{max}\{0,p-\gamma_{k}\}}^{p}\sum_{l=\left\lceil \frac{j}{q}\right\rceil}^{j}
\frac{(-1)^{l}}{l!}\Omega_{\gamma_{k}-p+j}^{k}D_{lq-j}^{l}c_{q}^{\frac{p-\gamma_{k}-lq}{q}}\Gamma\left(\frac{n+\gamma_{k}-p+lq}{q}\right)\;.
\end{equation}

The expansion in (\ref{68}) can now be used to obtain an expression for the sum in (\ref{59}) as follows
\begin{equation}\label{68c}
\frac{1}{(4\pi t)^{\frac{n}{2}}}\sum_{k=0}^{N+1}\left(\int_{\mathbb{R}^{n}}A_{k}(|\x|)e^{-tV(|\x|)}\diff\x\right)t^{k}=\frac{S_{n}}{c_{q}^{\frac{n}{q}}(4\pi)^{\frac{n}{2}}t^{\frac{n}{2}+\frac{n}{q}}}
\sum_{k=0}^{N+1}\left(\sum_{p=0}^{Mq+\gamma_{k}}T_{p}^{n,k}t^{\frac{p-\gamma_{k}}{q}+k}\right)+t^{N+1}O\left(t^{\frac{-n+Mq+1}{q}}\right)\;.
\end{equation}
From the right-hand-side of (\ref{68c}) we need to extract all the terms in the double-sum up to 
and including those of order $t^{N+1}$ and organize them in increasing powers of $t$. 
To correctly compute the coefficients of the small-$t$ expansion from (\ref{68c}) we need to take into account only the terms in the double-sum of (\ref{68c}) that have a power of $t$ satisfying the inequality $0\leq p-\gamma_{k}+qk\leq q(N+1)$. This automatically imposes a restriction on the parameter $M$ which needs to satisfy the equation $M=N+1$ when $k=0$ and $M=0$ when $k=N+1$.
Furthermore, in order to obtain the correct remainder in (\ref{68c}) one has to set, in the reminder term, $M=0$. This simply conveys the fact that the first term contributing to the reminder is the one in the inner sum for which $p=\gamma_{N+1}$.  
Under these conditions one can rearrange the double-sum in (\ref{68c}) to obtain
\begin{equation}\label{68cc}
\frac{1}{(4\pi t)^{\frac{n}{2}}}\sum_{k=0}^{N+1}\left(\int_{\mathbb{R}^{n}}A_{k}(|\x|)e^{-tV(|\x|)}\diff\x\right)t^{k}=\frac{S_{n}}{c_{q}^{\frac{n}{q}}(4\pi)^{\frac{n}{2}}t^{\frac{n}{2}+\frac{n}{q}}}\sum_{j=0}^{q(N+1)}a_{j}t^{\frac{j}{q}}
+O\left(t^{-\frac{n}{2}-\frac{n-1-q(N+1)}{q}}\right)\;,
\end{equation}
where the coefficients $a_{j}$ of the above asymptotic expansion can be obtained from the following relation by equating like powers 
of $t$
\begin{equation}\label{68d}
\sum_{k=0}^{N+1}\sum_{p=0}^{Mq+\gamma_{k}}T_{p}^{n,k}t^{\frac{p-\gamma_{k}+qk}{q}}=\sum_{j=0}^{q(N+1)}a_{j}t^{\frac{j}{q}}\;.
\end{equation}
In more details one can find
\begin{equation}\label{68e}
a_{j}=\sum_{v=0}^{v_{\textrm{max}}}T_{j+\gamma_{v}-vq}^{n,v}\;,
\end{equation}
where $v_{\textrm{max}}$ represents the largest integer satisfying the inequality $v_{\textrm{max}}q-\gamma_{v_{\textrm{max}}}\leq j$.

We need to analyze, now, the remainder term in (\ref{59}).
By recalling (\ref{58}) and by using spherical coordinates, $r_{N}(t,\x)$ in (\ref{59}) can be bounded, for small values of $t$, as follows
\begin{equation}\label{69}
\left|\int_{\mathbb{R}^{n}}r_{N}(t,\x)\diff\x\right|\leq S_{n}\delta_{N} t^{-\frac{n}{2}+N+2}\int_{0}^{\infty}r^{n-1}e^{-tV(r)}\diff r\;.
\end{equation}
The same change of variable exploited earlier, namely $c_{q}tr^{q}=s$, allows us to rewrite the integral in (\ref{69}) as  
\begin{equation}\label{70}
\int_{0}^{\infty}r^{n-1}e^{-tV(r)}\diff r=\frac{1}{qc_{q}^{\frac{n}{q}} t^{\frac{n}{q}}}\int_{0}^{\infty}s^{\frac{n}{q}-1}e^{-s}\exp\Bigg\{-t\sum_{j=0}^{q-1}c_{j}\left(\frac{s}{tc_{q}}\right)^{j/q}\Bigg\}\diff s\;.
\end{equation}
By using the result in (\ref{62c}) it is not very difficult to obtain the following small-$t$ asymptotic expansion 
of the integral in (\ref{70})
\begin{equation}\label{71}
\int_{0}^{\infty}r^{n-1}e^{-tV(r)}\diff r=\frac{1}{qc_{q}^{\frac{n}{q}} t^{\frac{n}{q}}}\sum_{j=0}^{Mq}t^{\frac{j}{q}}\int_{0}^{\infty}s^{\frac{n}{q}-1}e^{-s}H_{j}(s)\diff s
+O\left(t^{-\frac{n-Mq-1}{q}}\right)
\end{equation}  
From the above expression it is not difficult to realize that the leading term in the small-$t$ expansion of the integral 
is the one corresponding to the index $j=0$. This statement allows us to conclude that (\ref{69}) can be written as
\begin{equation}\label{72}
\left|\int_{\mathbb{R}^{n}}r_{N}(t,\x)\diff\x\right|\leq \delta_{N}\frac{S_{n}}{n}c_{q}^{-\frac{n}{q}} \Gamma\left(\frac{n}{q}+1\right) t^{-\frac{n}{2}-\frac{n}{q}+N+2}\;,
\end{equation} 
and, therefore,
\begin{equation}\label{72c}
\int_{\mathbb{R}^{n}}r_{N}(t,\x)\diff\x =O\left(t^{-\frac{n}{2}-\frac{n}{q}+N+2}\right)
\end{equation}

Finally, by using the results (\ref{68cc}) and (\ref{72c}) in (\ref{59}) we find the following small-$t$ asymptotic expansion of the trace of the Schr\"{o}dinger heat kernel
\begin{equation}\label{73}
K(t)=\frac{S_{n}}{c_{q}^{\frac{n}{q}}(4\pi)^{\frac{n}{2}}t^{\frac{n}{2}+\frac{n}{q}}}\sum_{j=0}^{N}a_{j}t^{\frac{j}{q}}
+O\left(t^{-\frac{n}{2}-\frac{n-1-N}{q}}\right)\;.
\end{equation}
A few remarks are in order at this point. It is well known that the leading small-$t$ behavior of the trace of the heat kernel 
$K(t)$ for a Laplace-type operator on a $d$-dimensional compact Riemannian manifold $\mathscr{M}$ with or without boundary is \cite{gilkey95,kirsten01}
\begin{equation}\label{73d}
K(t)\sim\frac{1}{(4\pi t)^{\frac{d}{2}}}\textrm{Vol}(\mathscr{M})\;,
\end{equation}
namely the leading term is $O(t^{-d/2})$. In the case of the Schr\"{o}dinger operator with an appropriate polynomial potential
studied here the form of the leading term of the trace of the associated heat kernel differs from the one described above. In fact, the expression (\ref{73}) clearly 
shows that the leading term of the small-$t$ asymptotic expansion of $K(t)$ not only depends on the dimension $n$ of the
Euclidean space but also on the degree of growth $q$ of the potential $V(r)$ as
\begin{equation}
K(t)\sim\frac{2^{1-d}\Gamma\left(\frac{d}{q}\right)}{qc_{q}^{\frac{d}{q}}\Gamma\left(\frac{d}{2}\right)t^{\frac{d}{2}+\frac{d}{q}}}\;,
\end{equation}
a behavior that was also observed in \cite{cognola06} once we set $c_{q}=1$. 
This implies, in particular, that it is the large-$r$ behavior of the potential $V(r)$ that determines the leading small-$t$ behavior of $K(t)$.

\section{Specific polynomial potentials}\label{sec5}

In this Section we use the general results obtained earlier to find the coefficients of the asymptotic expansion 
of the trace of the on-diagonal Schr\"{o}dinger heat kernel for specific spherically symmetric polynomial potentials defined on the Euclidean space $\mathbb{R}^{d}$.
According to the formula (\ref{68e}) the coefficients of the asymptotic expansion (\ref{73}) are written in terms of the expressions
$T_{p}^{n,k}$ in (\ref{68b}) which have been introduced in the process of organizing the various small-$t$ expansions, encountered 
throughout the calculations, in increasing powers of $t$.
The coefficients $T_{p}^{n,k}$ can be found with the help of a simple computer program once the dimension of the underlying Euclidean 
space and the polynomial potential have been specified. In addition, once a specific polynomial potential has been chosen, the 
coefficients $\Omega^{k}_{l}$ in (\ref{63}) can be extracted from the terms $A_{i}$ which, in turn, can be computed from the recurrence 
relation (\ref{23}). In particular one can find the following lower order coefficients (cf. \cite{cognola06,parker85}) 
\begin{eqnarray}\label{74a}
A_{0}(x)&=&1\;,\quad A_{1}(x)=0\;,\nonumber\\ 
A_{2}(x)&=&-\frac{1}{6}\Delta V(x)\;,\quad A_{3}(x)=-\frac{1}{60}\Delta^{2}V(x)+\frac{1}{12}\nabla_{j}V(x)\nabla^{j}V(x)\;,\nonumber\\
A_{4}(x)&=&-\frac{1}{840}\Delta^{3}V(x)+\frac{1}{72}(\Delta V(x))^2+\frac{1}{90}\nabla_{i}\nabla_{j}V(x)\nabla^{i}\nabla^{j}V(x)+\frac{1}{30}\nabla_{i}V(x)\nabla^{i}\Delta V(x)\;.
\end{eqnarray}
Higher order ones can be computed with an algebraic computer program.

The simplest and most studied example of polynomial potential is represented by a $d$-dimensional spherically symmetric harmonic oscillator potential 
$V(r)=cr^{2}$, with $c>0$. In this case the eigenvalues of the associated operator are known explicitly and they are 
$\lambda_{n_{1},\ldots,n_{d}}=\sqrt{c}(n_{1}+\cdots+n_{d}+d/2)$ with 
$n_{1},\ldots,n_{d}\geq 0$. The trace of the heat kernel can be computed in closed form and coincides with the partition function
of the $d$-dimensional harmonic oscillator, namely 
\begin{equation}\label{74}
K(t)=\frac{1}{\left[2\sinh\left(\sqrt{c}t\right)\right]^{d}}\;.
\end{equation} 
The small-$t$ asymptotic expansion for $K(t)$ can be easily obtained from its the Laurent expansion in the neighborhood of $t=0$. 
For instance, when $d=3$ we have from (\ref{74})
\begin{equation}\label{hk}
K(t)=\frac{1}{8c^{3/2}t^{3}}-\frac{1}{16\sqrt{c}t}+\frac{17\sqrt{c}t}{960}+O(t^{3})\;.
\end{equation}
The simple case of the spherically symmetric harmonic oscillator potential can provide a check of the method presented in this work. By setting $d=3$ and by using the 
potential $V(r)=cr^{2}$ in (\ref{74a}) it is not difficult to obtain 
\begin{equation}\label{75}
\Omega_{0}^{0}=1\;,\quad \Omega_{0}^{1}=0\;,
\end{equation}
\begin{equation}\label{75a}
\Omega_{0}^{2}=-c\;,\quad \Omega_{0}^{3}=0\;,\quad \Omega_{1}^{3}=0\;,\quad \Omega_{2}^{3}=\frac{c^{2}}{3}\;.
\end{equation}
In order to compute the first two non-vanishing coefficients of the small-$t$ expansion of $K(t)$ we need to set $q=2$ in (\ref{68b}) to get
\begin{equation}
T_{p}^{3,k}=\frac{1}{2}\sum_{j=\textrm{max}\{0,p-\gamma_{k}\}}^{p}\sum_{l=\left\lceil \frac{j}{2}\right\rceil}^{j}
\frac{(-1)^{l}}{l!}\Omega_{\gamma_{k}-p+j}^{k}D_{2l-j}^{l}c^{\frac{p-\gamma_{k}-2l}{2}}\Gamma\left(\frac{3+\gamma_{k}-p+2l}{2}\right)\;.
\end{equation} 
By noticing that $S_{3}=4\pi$, the last expressions together with (\ref{68d}) allow us to write the small-$t$ expansion of $K(t)$ in (\ref{73}), for the potential under consideration, as
\begin{equation}
K(t)=\frac{1}{2\sqrt{\pi}c^{3/2}t^3}\sum_{j=0}^{4}a_{j}t^{\frac{j}{2}}
+O\left(t^{-\frac{1}{2}}\right)\;,
\end{equation}
where the coefficients $a_{j}$ can be found to be
\begin{equation}
a_{0}=\frac{\sqrt{\pi}}{4}\;, \quad a_{1}=a_{2}=a_{3}=0\;,\quad a_{4}=-\frac{\sqrt{\pi}c}{8}\;,
\end{equation}
which agree, as expected, with the ones in (\ref{hk}).

For our next non-trivial example we consider a spherically symmetric quartic oscillator potential, namely a potential of the general form 
\begin{equation}\label{75}
V(r)=c_{0}+c_{2}r^{2}+c_{4}r^{4}\;,
\end{equation}
where we assume that $V(r)$ is defined on $\mathbb{R}^{3}$. Here and in the remaining examples below, we assume that the coefficients $c_{j}$ in the potential 
function $V(r)$ are such that $V(r)\to\infty$ as $r\to\infty$ and  $\textrm{inf}_{r>0}V(r)\geq 0$.  
By using (\ref{75}) in (\ref{74a}) one can obtain the following coefficients $\Omega_{i}^{j}$ 
\begin{equation}\label{76}
\Omega_{0}^{0}=1\;,\quad \Omega_{0}^{1}=0\;,
\end{equation}
\begin{equation}\label{76a}
\Omega_{0}^{2}=-c_{2}\;,\quad \Omega_{1}^{2}=0\;,\quad \Omega_{2}^{2}=-\frac{10}{3}c_{4}\;,
\end{equation}
\begin{equation}\label{76b}
\Omega_{0}^{3}=-2c_{4}\;,\quad \Omega_{2}^{3}=\frac{1}{3}c_{2}^{2}\;,\quad \Omega_{4}^{3}=\frac{4}{3}c_{2}c_{4}\;,\quad \Omega_{6}^{3}=\frac{4}{3}c_{4}^{2}\;, 
\quad \Omega_{1}^{3}=\Omega_{3}^{3}=\Omega_{5}^{3}=0\;.
\end{equation}
The terms $\Omega_{i}^{j}$ provided above can be used to obtain the coefficients of the small-$t$ asymptotic expansion of the trace of the 
heat kernel up to and including the one proportional to $t^{1/4}$. By setting $q=4$ in (\ref{68b}) we get 
the expression 
\begin{equation}\label{77}
T_{p}^{3,k}=\frac{1}{4}\sum_{j=\textrm{max}\{0,p-\gamma_{k}\}}^{p}\sum_{l=\left\lceil \frac{j}{4}\right\rceil}^{j}
\frac{(-1)^{l}}{l!}\Omega_{\gamma_{k}-p+j}^{k}D_{4l-j}^{l}c_{4}^{\frac{p-\gamma_{k}-4l}{4}}\Gamma\left(\frac{3+\gamma_{k}-p+4l}{4}\right)\;,
\end{equation}
which together with (\ref{68e}) allows us to compute the coefficients $a_{j}$. In more details the asymptotic expansion of the trace of the heat kernel 
when a quartic oscillator potential in considered has the form
\begin{equation}\label{78}
K(t)=\frac{1}{2\sqrt{\pi}c_{4}^{3/4}t^{9/4}}\sum_{j=0}^{10}a_{j}t^{\frac{j}{4}}+O(t^{1/2})\;,
\end{equation}
where the non-vanishing coefficients $a_{j}$ have the form
\begin{equation}\label{80}
a_{0}=\frac{1}{4}\Gamma\left(\frac{3}{4}\right)\;,\quad a_{2}=-\frac{c_{2}}{16\sqrt{c_{4}}}\Gamma\left(\frac{1}{4}\right)\;,
\end{equation}
\begin{equation}\label{81}
a_{4}=\frac{1}{c_{4}}\left(\frac{3}{32}c_{2}^{2}-\frac{1}{4}c_{0}c_{4}\right)\Gamma\left(\frac{3}{4}\right)\;,
\end{equation}
\begin{equation}\label{82}
a_{6}=-\frac{1}{c_{4}^{3/2}}\left(\frac{5}{384}c_{2}^{3}-\frac{1}{16}c_{0}c_{2}c_{4}+\frac{5}{48}c_{4}^{2}\right)\Gamma\left(\frac{1}{4}\right)\;,
\end{equation}
\begin{equation}\label{83}
a_{8}=\frac{1}{c_{4}^{2}}\left(\frac{7}{512}c_{2}^{4}-\frac{3}{32}c_{0}c_{2}^{2}c_{4}+\frac{3}{16}c_{2}c_{4}^{2}+\frac{1}{8}c_{0}^{2}c_{4}^{2}\right)
\Gamma\left(\frac{3}{4}\right)\;,
\end{equation}
and
\begin{equation}\label{84}
a_{10}=-\frac{1}{c_{4}^{5/2}}\left(\frac{3}{2048}c_{2}^{5}-\frac{5}{384}c_{0}c_{2}^{3}c_{4}+\frac{13}{384}c_{2}^{2}c_{4}^{2}+\frac{1}{32}c_{0}^{2}c_{2}c_{4}^{2}
-\frac{5}{48}c_{0}c_{4}^{3}\right)\Gamma\left(\frac{1}{4}\right)\;.
\end{equation}

As a final example we consider a spherically symmetric sestic oscillator potential
\begin{equation}\label{85}
V(r)=c_{0}+c_{2}r^{2}+c_{4}r^{4}+c_{6}r^{6}\;,
\end{equation}
defined on $\mathbb{R}^{3}$. The coefficients $\Omega_{i}^{j}$ associated with the above potential 
are
\begin{equation}\label{86}
\Omega_{0}^{0}=1\;,\quad \Omega_{0}^{1}=0\;,
\end{equation}
\begin{equation}\label{87}
\Omega_{0}^{2}=-c_{1}\;,\quad \Omega_{2}^{2}=-\frac{10}{3}c_{2}\;,\quad \Omega_{4}^{2}=-7c_{3}\;,\quad \Omega_{1}^{2}=\Omega_{3}^{2}=0\;,
\end{equation}
\begin{eqnarray}\label{88}
\Omega_{0}^{3}&=&-2c_{2}\;,\quad \Omega_{2}^{3}=-14c_{3}+\frac{1}{3}c_{1}^{2}\;,\quad \Omega_{4}^{3}=\frac{4}{3}c_{1}c_{2}\;,\quad \Omega_{6}^{3}=\frac{4}{3}c_{2}^{2}+2c_{1}c_{3}\\
\Omega_{8}^{3}&=&4c_{2}c_{3}\;,\quad \Omega_{10}^{3}=3c_{3}^{2}\;,\quad \Omega_{1}^{3}=\Omega_{3}^{3}=\Omega_{5}^{3}=\Omega_{7}^{3}=\Omega_{9}^{3}=0
\end{eqnarray}
which have been computed by using (\ref{75}) in (\ref{74a}). These terms can be used to compute the asymptotic expansion of the trace of the heat kernel up to and including 
the terms proportional to $t^{-1/3}$. Setting $q=6$ in (\ref{68b}) provides the formula
\begin{equation}\label{89}
T_{p}^{3,k}=\frac{1}{6}\sum_{j=\textrm{max}\{0,p-\gamma_{k}\}}^{p}\sum_{l=\left\lceil \frac{j}{6}\right\rceil}^{j}
\frac{(-1)^{l}}{l!}\Omega_{\gamma_{k}-p+j}^{k}D_{6l-j}^{l}c_{6}^{\frac{p-\gamma_{k}-6l}{6}}\Gamma\left(\frac{3+\gamma_{k}-p+6l}{6}\right)\;,
\end{equation}
which used in conjunction with (\ref{68e}) will produce the coefficients $a_{j}$. More explicitly, the small-$t$ asymptotic expansion of the trace of the heat kernel associated
with the Schr\"{o}dinger operator endowed with a sestic oscillator potential of the form (\ref{85}) reads
\begin{equation}\label{90}
K(t)=\frac{\sqrt{\pi}}{2\sqrt{c_{6}}t^{2}}\sum_{j=0}^{10}a_{j}t^{\frac{j}{6}}+O(t^{-1/6})\;,
\end{equation} 
with the non-vanishing coefficients 
\begin{equation}\label{92}
a_{0}=\frac{\sqrt{\pi}}{6}\;,\quad a_{2}=-\frac{c_{4}}{36c_{6}^{2/3}}\Gamma\left(\frac{1}{6}\right)\;,
\end{equation}
\begin{equation}
a_{4}=\frac{1}{c_{6}^{4/3}}\left(\frac{5}{72}c_{4}^{2}-\frac{1}{6}c_{2}c_{6}\right)\Gamma\left(\frac{5}{6}\right)\;,
\end{equation}
\begin{equation}
a_{6}=-\frac{\sqrt{\pi}}{c_{6}^{2}}\left(\frac{1}{48}c_{4}^{3}-\frac{1}{12}c_{2}c_{4}c_{6}+\frac{1}{6}c_{0}c_{6}^{2}\right)\;,
\end{equation}
\begin{equation}
a_{8}=\frac{1}{c_{6}^{8/3}}\left(\frac{91}{31104}c_{4}^{4}-\frac{7}{432}c_{2}c_{4}^{2}c_{6}+\frac{1}{36}c_{0}c_{4}c_{6}^{2}+\frac{1}{72}c_{2}^{2}c_{6}^{2}-\frac{7}{72}c_{6}^{3}\right)\Gamma\left(\frac{1}{6}\right)\;,
\end{equation}
and
\begin{equation}
a_{10}=-\frac{1}{c_{6}^{10/3}}\left(\frac{187}{31104}c_{4}^{5}-\frac{55}{1296}c_{2}c_{4}^{3}c_{6}+\frac{5}{72}c_{2}^{2}c_{4}c_{6}^{2}
+\frac{5}{72}c_{0}c_{4}^{2}c_{6}^{2}-\frac{1}{6}c_{0}c_{2}c_{6}^{3}-\frac{5}{24}c_{4}c_{6}^{3}\right)\Gamma\left(\frac{5}{6}\right)\;.
\end{equation}
We would like to point out that in all previous examples we have only computed, for the sake of brevity, the first few non-vanishing coefficients of the small-$t$ asymptotic expansion of $K(t)$. Obviously higher order coefficients can be computed easily once higher order coefficients $A_{k}$ are found.

\section{Conclusions}

In this work we have shown that the small-$t$ asymptotic expansion of the 
trace of the heat kernel for a Laplace operator endowed with a spherically symmetric polynomial potential
can be obtained by termwise integration of the small-$t$ expansion of the associated on-diagonal heat kernel.
In order for the ensuing integrals to be well defined we used a resummed form of the heat kernel which 
contains the negative exponent of the potential (cf. (\ref{24a})). 
The method outlined in Section \ref{sec4} has been implemented in Section \ref{sec5} to find the small-$t$ asymptotic expansion
of the trace of the Schr\"{o}dinger heat kernel $K(t)$ for some specific spherically symmetric polynomial potentials.
The approach developed in this work provided, in a fairly straightforward way, explicit formulas for the terms of the asymptotic expansion of $K(t)$.

This work represents only a first step towards the development of a more general technique that would allow the explicit computation of 
the coefficients of the asymptotic expansion of the trace of the heat kernel for suitable Laplace-type operators on unbounded domains. 
The results obtained here could be useful for the study of quantum fields that are confined within polynomial potentials. 
In particular our paper could complement the analysis of Bose-Einstein condensation in polynomial potentials which 
has been considered, for example, in \cite{bagnato,kirsten98}.      

While this work is focused on spherically symmetric polynomial potentials, it would certainly be very interesting to study and classify 
all the types of potentials that allow for a termwise integration of the resummed form of the expansion of the heat kernel, hence extending and 
improving the results obtained in this paper. 
Another important aspect of the study of heat kernels on unbounded domains, consists in the analysis of the heat kernel asymptotic expansion for a Laplace operator with a 
spherically symmetric and exponentially increasing potential. In this case the resulting Schr\"{o}dinger operator is essentially selfadjoint
with a real spectrum which is increasing and bounded from below. It would be of particular interest to analyze whether 
the method described in this work can be extended to the case of exponentially increasing potentials. It has been shown in \cite{cognola06} that 
the small-$t$ expansion of $K(t)$ in the presence of exponentially increasing potentials is non-standard and contains 
logarithmic terms similar to the case of non-smooth manifolds (see e.g. \cite{bordag96a,flachi10,fucci12}).
Due to this non-standard behavior of the small-$t$ asymptotic expansion we expect that the termwise integration method developed in this work will need 
to be somewhat modified in order to treat the case of exponentially increasing potentials. Based on the formulas developed in this paper one cannot  
infer, in a simple and direct way, how the logarithmic terms appear in the small-$t$ asymptotic expansion of the trace of the heat kernel in the case of exponentially increasing potentials. However, if the termwise integration method can be shown to be valid in this case then the logarithmic term
in the small-$t$ asymptotic expansion arises from the integration of the $A_{0}$ term of the small-$t$ expansion, in its partially resummed form, of $K(t|x,x)$ in (\ref{7})
as it has been argued in \cite{cognola06}.

\end{document}